\documentclass[twocolumn,aps,prl,amsmath,showpacs]{revtex4-1}
\usepackage{graphicx,dcolumn,bm,amssymb,amsmath}
\usepackage{color}

\def\be{\begin{equation}}
\def\ee{\end{equation}}

\begin{document}
\author{A. Zaccone$^{1,3}$, P. Schall$^{2}$, and E. M. Terentjev$^{3}$}
\affiliation{${}^1$Physics Department and Institute for Advanced Study, Technische Universit\"{a}t M\"{u}nchen,
85748 Garching, Germany}
\affiliation{${}^2$Van der Waals-Zeeman Institute, University of Amsterdam, The Netherlands}
\affiliation{${}^3$Cavendish Laboratory, JJ Thomson Avenue, CB30HE Cambridge,
U.K.}

\begin{abstract}
\noindent The atomic theory of elasticity of amorphous solids, based on the
nonaffine response formalism, is extended into the nonlinear stress-strain
regime by coupling with the underlying irreversible many-body dynamics. The
latter is implemented in compact analytical form using a qualitative method for
the many-body Smoluchowski equation. The resulting nonlinear stress-strain
(constitutive) relation is very simple, with few fitting parameters,
yet contains all the microscopic physics. The theory is successfully tested
against experimental data on metallic glasses, and it is able to reproduce the
ubiquitous feature of stress-strain overshoot upon varying temperature and
shear rate. A clear atomic-level interpretation is provided for the stress
overshoot, in terms of the competition between the elastic instability caused
by nonaffine deformation of the glassy cage and the stress buildup due to
viscous dissipation.
\end{abstract}

\title{Microscopic origin of nonlinear non-affine deformation and stress overshoot in bulk metallic
glasses}
\maketitle

The microscopic mechanism controlling the nonlinear deformation behavior and
plasticity of crystals has been rationalized in terms of dislocation mobility
starting with the seminal contributions of Orowan~\cite{Orowan},
Polanyi~\cite{Polanyi}, and G. I. Taylor~\cite{Taylor}, all in 1934. These were
followed by mathematically more refined treatments and advances in dislocation
dynamics, among others, by Peierls and Nabarro~\cite{Nabarro}. Jointly with the
atomic theory of linear elasticity developed by Born and coworkers~\cite{Born},
the understanding of both linear and nonlinear deformations of crystals has
reached an advanced level down to the atomic-scale, with many 
applications in metallurgy.

In contrast, the deformation behavior of amorphous solids (e.g. glasses), which
lack both orientational and translational symmetry, has remained more elusive.
The lack of local centers of inversion symmetry makes the Born-Huang affine
approximation for down-scaling the macroscopic deformation at the atomic level
inapplicable~\cite{Alexander}. Only recently the non-affine deformation
formalism has brought a deeper understanding of atomic-scale deformation in the
linear elastic regime~\cite{Lemaitre,Head, Zaccone2011,Yoshino}. In the absence
of a local center of inversion symmetry, as is the case in glasses, the forces
transmitted upon deformation by the neighbors on any atom do not balance,
and so require additional displacements (called non-affine) in order
to be locally equilibrated.

In the homogeneous nonlinear deformation regime of amorphous solids, the transition to
plastic behavior is also problematic. The usual concept of dislocation glide or
climb, which proved so useful in describing the crystal plasticity, is
difficult to apply when no long-range order exists and one cannot identify
defects that could mediate the plastic flow~\cite{Edwards}. Instead, the local shear transformation zones (STZs), where concentrated rearrangements of atoms or particles occur, have been identified as carriers of the plastic flow in amorphous solids~\cite{ArgonKuo79,Schall07,Langer}. Such STZs exhibit similar long-range stress fields as dislocation dipoles~\cite{Eshelby}, and they have been shown to form preferentially at structurally weak spots of the material~\cite{ManningLiu}.

In spite of these efforts, fundamental points remain unclear, including the
actual topology of STZs, which is not very well defined, unlike
dislocations in crystal. Also, it is not clear how STZs relate to the underlying non-affine displacements, which are intrinsic to disordered solids, are known to strongly affect the elastic deformation at the linear level and may
contribute to the overall vanishing of shear rigidity.
Most importantly, a simple atomic-scale picture of the transition from the
elastic non-affine deformation to flow, mediated by the amorphous structure, is
currently lacking.

Here we propose such a microscopic mechanism, following a different route.
Unlike previous approaches, we start from the non-affine linear response and
then couple it to the irreversible shear-induced many-body dynamics causing
structural rearrangements of the glassy cage, and the stress non-linearity.
The resulting theory has the advantage of being simple and fully analytical, as
opposed to earlier more involved approaches that rely on hardly testable assumptions. 
Despite its simplicity, our model can accurately reproduce the
stress overshoot~\cite{Johnson,Inoue} of metallic glasses in fairly good agreement
with experiments. Further, it provides the fundamental connection between
non-affine deformation, local cage rearrangements and plastic creep,
and suggests a more microscopic interpretation of STZs in terms of local
connectivity and microstructural heterogeneity.

The starting point of our analysis is the free energy of deformation of
disordered solids which can be written as
$F=F_A(\gamma)-F_{NA}(\gamma)$, with two distinct
contributions arising in response to the macroscopic shear deformation $\gamma$.
The first, $F_{A}$, is the standard affine deformation energy as one finds in
the Born-Huang theory of lattice dynamics~\cite{Born}. Affinity means that
every particle follows the macroscopic shear, and the
associated interatomic displacements are simply proportional to $\gamma$. The
non-affine contribution, $F_{NA}$, lowers the free energy of deformation due to
additional non-affine displacements~\cite{Zaccone2011,Zaccone2013}. In a
nutshell, if the particles are not local centers of lattice symmetry, there is
an imbalance of forces on every particle when the deformation is applied,
unlike in crystals with inversion symmetry. This additional net force acting on
every particle in disordered solids has to be relaxed through additional
(non-affine) motions that occur on top of the affine displacements dictated by
the macroscopic strain. The non-affine displacements perform internal work
against the potential-field of the solid, which results in a net negative
contribution to the free energy of deformation, reducing the effect of the
basic affine elastic energy.  As shown in earlier work~\cite{Zaccone2013}, if
the interatomic forces are purely central, with $\kappa$ the spring constant of
a harmonic bond and $R_{0}$ the equilibrium distance between nearest-neighbors,
and if $\phi$ is the
atomic packing fraction in the solid, the shear modulus can be written as
$G=\frac{2}{5\pi}(\kappa\phi/R_{0})(n_{b}-n_{b}^{c})$. Here $n_{b}$ denotes the
average
coordination number of mechanical bonds per atom (a more precise definition is
explored below),
while the critical coordination number $n_{b}^{c}$ is a result of non-affine
adjustments.
For purely central bond potential, $n_b^c=2d$, where $d$ is the space
dimension. The situation is slightly different with covalently-bonded glasses
(with
non-central forces) where typically $n_{b}^{c} \approx 2.4$ is set by the atomic valency, and the
coordination at rest $n_{b}^{0}$ is much lower than 12.
For closely packed materials like metals, it is useful to refer to the effective potential
of mean local force $V_{\mathrm{eff}}$. This is a standard concept in statistical mechanics,
defined in equilibrium as $V_{\mathrm{eff}}/kT=-\ln g(r)$. The attractive
minimum in $V_{\mathrm{eff}}$ is located at the same interatomic distance
$R_{0}$ at which $g(r)$ exhibits its first peak. In this way,
$V_{\mathrm{eff}}$ effectively accounts for complex many-body effects on the
pair interaction, and can be described using the pseudopotential theory of
metals. The connectivity at zero-shear can thus be inferred from the knowledge
of $V_{\mathrm{eff}}$: it is reasonable that only those atoms which are within a distance
$R_{0}$ (i.e. within the attractive minimum) from the given atom contribute to
$n_{b}$. 

Now consider that the number of
interatomic bonds may change during the deformation due to shear-induced
distortion of the ``cage'' formed by neighbors surrounding a given particle.
This leads to a dependence $n_{b}(\gamma)$ in the previous expression
for $G$, which makes $G$ vary with strain. The cage dynamics is governed by the many-body Smoluchowski equation
with an added shear-force term $\propto \kappa
R_{0}\gamma$~\cite{Fisher,Chandra,Dhont}. The equation can be written for the
radial distribution function $g(r)$~\cite{Fisher}, whose first peak then
depends on the shear strain $\gamma$, and thus controls the variation of
$n_{b}$ with strain~\cite{Dhont}. For a quasistatic deformation, no shear-rate
and time dependencies need be considered, and the general spherically-averaged
form of the steady-state expression takes the form
$g(r,\gamma)=\exp[-V_{\mathrm{eff}}/kT+h(r)\gamma]$, where $h(r)$ is a suitable
decaying function of $r$ (see~\cite{Dhont} for detail). 

\begin{figure}
\centering
{\includegraphics[width=0.85\columnwidth]{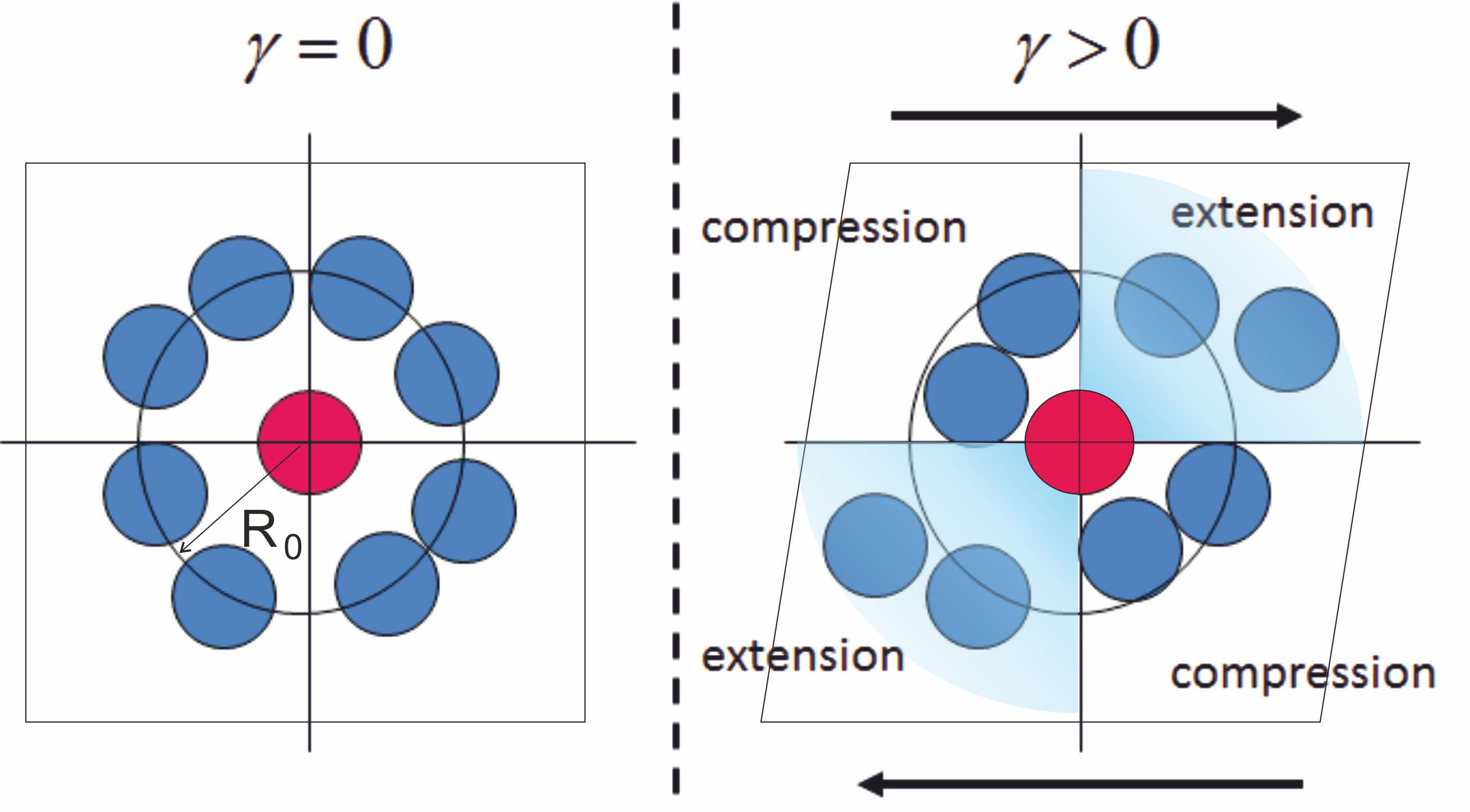}
\caption{(Color online) A cage-breaking model: Without shear, the number
of particles moving in and out the cage is equal. In the presence of shear
$\gamma$, the
number of particles moving out of the cage in the sectors of local extension
axis is higher than in the sectors of the compression axis. }
\label{fig1}}
\end{figure}

Consider a scheme of local deformation around a given particle (atom),
Fig.\ref{fig1}. In the extension sectors of solid angle under shear, the
neighbors are pulled farther apart from the test atom at the center of the
cage. As the neighbors cross the $R_{0}$ boundary in the outward direction,
under the action of shear in the extension sectors, they cease to contribute to
$n_{b}$. In the compression sectors, atoms are pushed inwards by the local
deformation field, which could lead to the formation of new mechanical contacts
with atoms which were previously just outside the $R_{0}$ limit. However, this latter
effect must be strongly opposed by the excluded-volume interactions between
atoms, which limits the formation of new contacts, while the soft
attraction for the departing atoms has no such constraint.
Hence, the shear-induced depletion of mechanical bonds in the extension sectors
cannot be exactly compensated by formation of new bonds in the compression
sectors. This results in a net decrease of the first peak of the
spherically-averaged correlation function $g(r,\gamma)$, which implies $h(r)<0$
in the solution to the many-body Smoluchowski equation. Further, if the
deformation is applied at a finite rate, as is the case for a strain ramp
$\dot{\gamma}=\gamma/t$, {the solution to the governing time-dependent
Smoluchowski equation for the pair distribution function is formally identical
to the solution to the time-dependent Schroedinger equation in a transformed
effective potential~\cite{risken}. The general solution can thus be written as
a superposition of steady-state eigenfuctions $\phi_{k}(r)$, where $k$ labels
the energy-level
of the eigenfunction. The time-dependent part is expressed as usual in terms of
the eigenvalues $\lambda_{k}$, giving
the general form:}
$g(r,t)=\sum_{k} {{\phi _k}} (r){e^{ - {\lambda _k}t}}$.
At lower deformation rates, the sum is dominated by the
lowest non-zero eigenvalue, which in this case is the
inverse of the cage relaxation time, $\lambda_{1}=1/\tau_{c}$. Hence,
$g(R_{0}) \sim e^{-\gamma/\dot{\gamma} \tau_{c}}$. 

Recall the definition of coordination number in amorphous systems,
$n_{b}=4\pi\rho\int_{peak}g(r)r^{2}dr$, with $\rho$ the mean density
\cite{Fisher,Hansen}. Evidently,
$n_{b}$ has roughly the same (time) dependence on $\gamma$, $\dot{\gamma}$, and
$\tau_{c}$ as does $g(R_{0})$. In the limit $\gamma \gg 1$,
all the mechanical neighbors must have been pealed off from the extension
sectors, while the neighbors of the compression sectors have remained on
average in their original positions, being pushed inwards continuously by the
action of shear. This implies $n_{b}\rightarrow n_{b}^{0}/2$ as
$\gamma \gg 1$, where $n_{b}^{0}=12$ is the equilibrium coordination number of
most metallic glasses at rest~\cite{Egami}. This recovers fluid
behavior at large strain, in accordance with the marginal stability
principle~\cite{Thorpe}: $G\propto[(n_{b}^{0}/2)-6]=0$ at
$\gamma \gg 1$, in 3D.

A simple, general expression for the evolution of $n_{b}$, which contains the mechanism depicted in Fig.1 for the net change in coordination number due to thermal motion and 
shear-induced distortion is as follows:
\begin{equation}\label{1}
n_{b}(\gamma)=\frac{n_{b}^{0}}{2}(1+e^{-A\gamma}),
~~~ \mathrm{with}~~A=\frac{\Delta}{k_BT}+\frac{1}{\dot{\gamma}\tau_{c}}.
\end{equation}
This expression is also consistent with the qualitative behavior $g(R_{0}) \sim e^{-\gamma/\dot{\gamma} \tau_{c}}$ which results from the Smoluchowski dynamics (shear-induced exponential depletion of neighbors in the extension sector upon increasing strain), and it complies with the limits expected based on marginal stability analysis. The latter means that Eq.(1) explicitly
recovers $G=0$ in the limit $\gamma \gg \dot{\gamma}\tau_{c}$, when the cage is
emptied in the two extension sectors. $\Delta$ represents an energy barrier for
the shear-induced breaking of the cage, which is related to the energetics of
thermal cage-breaking, hence to the glass transition. Assuming that the cage
melts at the glass transition
temperature $T_{g}$, we then have the approximate relation $\Delta=k_{B}T_{g}$.
Inserting the expression for $n_{b}(\gamma)$ in the
the free energy of deformation
$F_{el}=\frac{1}{2}K[n_{b}(\gamma)-n_{b}^{c}]\gamma^{2}$, and differentiating,
we obtain the
nonlinear elastic stress-strain relationship for the metallic glass:
\begin{equation}\label{2}
\sigma_\mathrm{el}(\gamma) = \frac{1}{4} n_{b}^{0} K\gamma \cdot
e^{-\gamma \left(\frac{T_{g}}{T}+\frac{1}{\dot{\gamma}\tau_{c}}\right)}
\left[2-\gamma\left(\frac{T_{g}}{T}+
\frac{1}{\dot{\gamma}\tau_{c}}\right)\right],
\end{equation}
with the shorthand $K=\frac{2}{5\pi}(\kappa\phi/R_{0})$. As one can easily
check, this expression features an elastic instability corresponding to a point
of maximum stress in the stress-strain curve, see Fig. \ref{fig2} below. At this point
$G(\gamma)=0$ because the elastic energy associated with the bonds that
survived the shear-induced cage-breaking process, is no longer enough to
compensate the lattice deformation energy lost to non-affine motions (in other
words, $n_{b}(\gamma)-n_{b}^{c}=0$).

To complete the picture, it is necessary to also consider the viscous
contribution to the total stress. It is known that for deformations that are
not quasistatic, i.e. with $\dot{\gamma}>0$, microscopic friction induces a
resistance to the atomic displacements, even in perfect crystals~\cite{Landau}.
The microscopic friction is associated with a viscosity $\eta$, and a viscous
(Maxwell) relaxation time $\tau_v$. For a constant rate of strain, this stress
contribution can be written in terms of the relaxation modulus as
$\sigma(t)=\dot{\gamma} \int_0^t G(s)ds$. For the linear viscoelastic solid (Zener solid), the relaxation modulus
is given as $G(t)=G + G_{R}\exp[-t/\tau_v]$ \cite{Zener}, where
$\tau_v=\eta/G_{R}$ and $G_{R}=G_{0}-G$, where $G_{0}$ is the instantaneous
(infinite-frequency) shear modulus.
The total stress follows upon integration as
$\sigma_{\mathrm{tot}}=\sigma_\mathrm{el}+\sigma'$, where the viscous addition is
$\sigma'=\dot{\gamma} G_{R} \tau_v (1-e^{-\gamma/\dot{\gamma}\tau_v})$, while
the elastic part is given by the Eq. (\ref{2}), leading to:
\begin{equation}\label{3}
\sigma_\mathrm{tot}=\sigma_\mathrm{el}(\gamma) + \eta \dot{\gamma} \cdot
(1-e^{-\gamma/\dot{\gamma}\tau_{v}}).
\end{equation}
In a compact form, this equation contains all the relevant atomic-level
physics: interatomic pseudopotential (contained in $K$), non-affine
displacements (showing in the negative $-n_b^c$), shear-induced changes in
local atomic connectivity $n_b(\gamma)$ also including the
thermally-activated cage-distortion, and the viscous dissipation due to the
microscopic friction. The expression recovers the elastic limit at small
strain, where $\sigma_{\mathrm{tot}} \approx n_b^0 K \gamma$, and in the
opposite
limit of $\gamma \gg 1$ it recovers plastic flow, $\sigma_{\mathrm{tot}}
\rightarrow \eta \dot{\gamma}$. By taking the first derivative of Eq.(3) with
respect to $\gamma$ and setting it to zero, the yield strain $\gamma_{y}$ (or
the strain at which the stress is maximum, at the top of the overshoot: see
Figs. \ref{fig2} and \ref{fig3}) can be evaluated. Two non-dimensional
parameters control the outcome: $H=4\eta/(12K\tau_{v})$ and
$B=A\dot{\gamma}\tau_{v}$. A general solution of the resulting transcendent
equation cannot be found, but an approximate analysis is possible. Whenever the
condition $H\gg B$ is satisfied (which is mostly the case in
practice), the following relation for the yield strain holds:
\begin{equation}
\gamma_{y} \approx \frac{0.6}{T_{g}/{T}+{1}/{\dot{\gamma}\tau_{c}}}.
\end{equation}
The yield strain is thus an increasing function of both $T$ and $\dot{\gamma}$.

We shall now test how this theory performs in comparison with experimental
data. The mechanical response of metallic glasses has been studied extensively.
When the response is not affected by shear banding, i.e. at not too high shear
rates, the stress-strain relation typically features an overshoot with a
maximum in the stress beyond which the yielding regime sets in. This eventually
transforms into the viscous Newtonian flow in the large strain limit. This overshoot 
behaviour provides a benchmark for theories of deformation: the extent of the
overshoot is modulated in a nontrivial manner by temperature and shear-rate.
Here for our comparison we use the classical experiments done by the Johnson
group~\cite{Johnson} on the commercial amorphous alloy
$\mathrm{Zr}_{41.2}\mathrm{Ti}_{13.8}\mathrm{Cu}_{12.5}
\mathrm{Ni}_{10}\mathrm{Be}_{22.5}$. In these experiments, the tensile stress
was measured. This is directly proportional to and controlled by the total
shear stress given by our theory, since the Poisson ratio does not vary much
over the strain window under consideration. This is of course an uncontrolled
approximation, but it certainly cannot change the qualitative comparison
appreciably, given the very narrow range within which the Poisson ratio is
allowed to vary.

The comparison between predictions of the theory and experimental data
\cite{Johnson}, is shown in Figs. \ref{fig2} and \ref{fig3}. The
non-trivial fitting parameters required by our theory are the two relaxation times:
the cage
relaxation time $\tau_{c}$, and the viscous relaxation time $\tau_v$. The theory
is able to reproduce the experimental data rather accurately, in spite of the
mathematical simplicity of Eqs. (\ref{2}) and (\ref{3}). In particular, the
theory captures
the effects of varying the temperature and the shear rate on the emergence and
extent of the overshoot.

\begin{figure} [h]
\includegraphics[width=.45\textwidth]{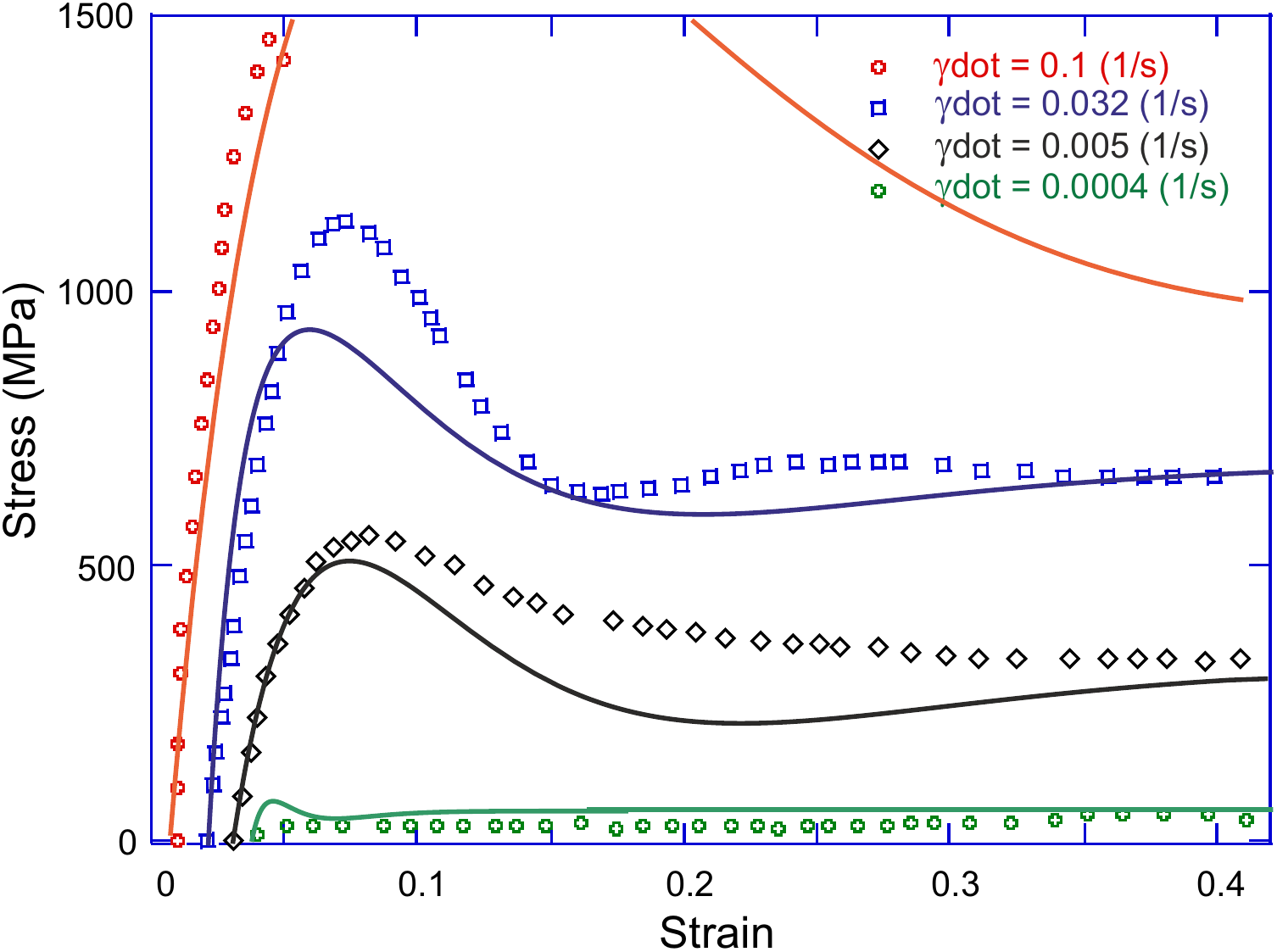}
\caption{Comparison between theoretical predictions and experimental data: (a)
Eq.(3) with $\dot{\gamma}=0.1\mathrm{s}^{-1}$ and
$T_g=625$K for all curves. $K$ and $\tau_v$ were chosen to match the
experimental data in the elastic and viscous-flow regimes, respectively. List
of all parameter values is given in Table I.  In
(b), the plot of data from \cite{Johnson}, with the curves artificially shifted
to the right to avoid overlapping. }
\label{fig2}
\end{figure}

The existence and the amplitude of the overshoot are due to the competition
between the elastic instability driven by non-affine shear-induced cage
breakup and the build-up of viscous stress, respectively. In particular, when
the elastic
instability sets in, it causes the stress to go through a maximum value
$\sigma_{\mathrm{max}}$ and to subsequently decrease with further increasing
strain, whereas the viscous contribution $\sigma'$ increases monotonically up
to the final Newtonian plateau. This is evident from Eq. (\ref{3}). The maximum
stress is directly controlled by the local atomic connectivity $n_{b}$
decreasing with $\gamma$, a process controlled by the cage-breaking relaxation
time $\tau_{c}$ and the activation energy represented as $T_g/T$. Both of them,
in turn, control
the critical strain $\gamma_{y}$ (yield point) associated with the maximum
stress. Hence they
also control the magnitude of the maximum stress at the yield point. 

Increasing
$T$ at fixed $\dot{\gamma}$ has the effect of making the cage more easily breakable, equivalent to a
lower activation energy $\Delta$, leading to a lower yield strain $\gamma_{y}$.
Therefore, the maximum stress which can be reached must decrease upon
increasing $T$  (at fixed strain rate $\dot{\gamma}$).

\begin{figure} [h]
\includegraphics[width=.45\textwidth]{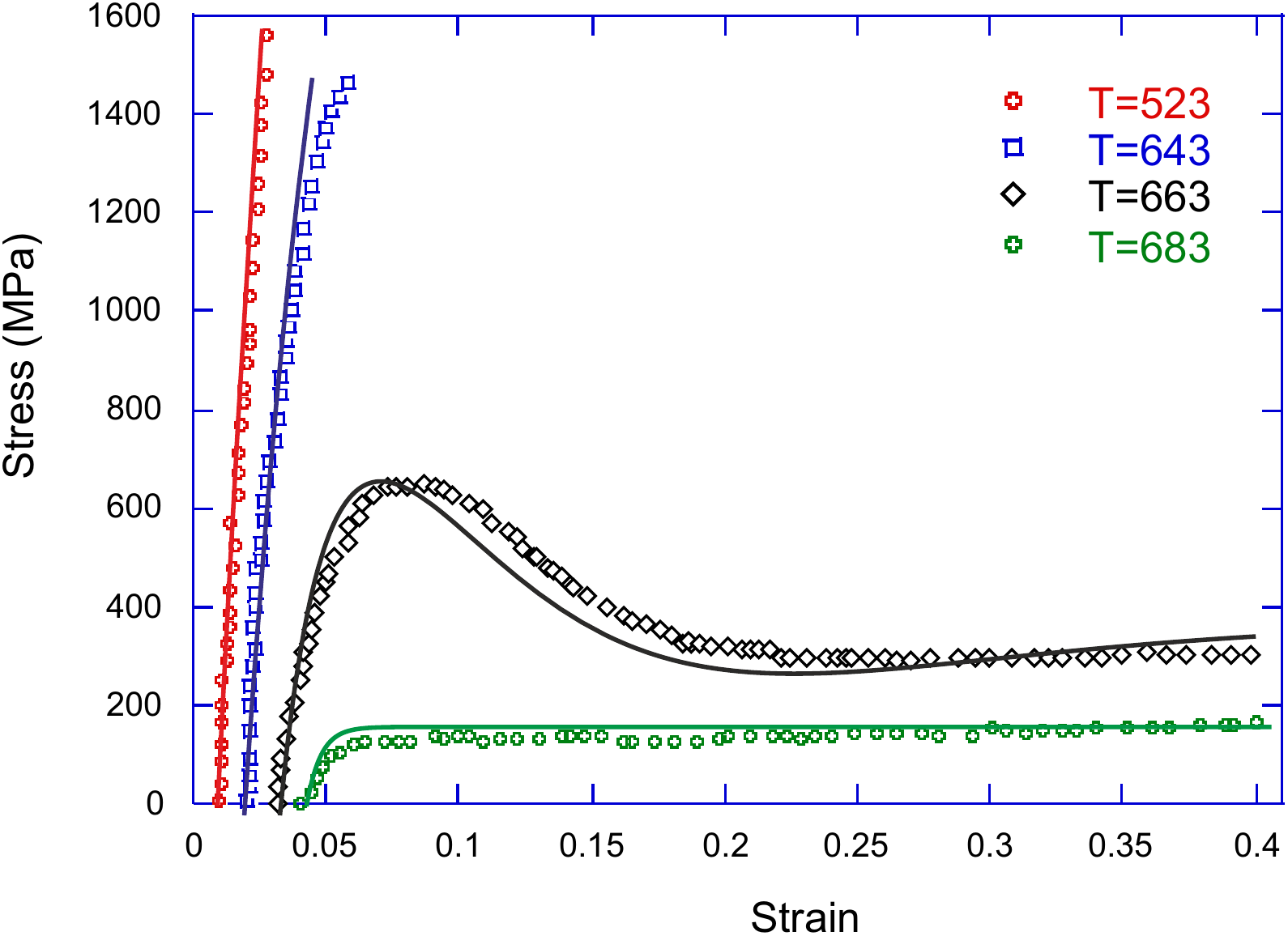}
\caption{Comparison between theoretical predictions and experimental data: (a)
Eq.(3) with a constant $T=643$K and $T_g=625$K for all
curves. $K$ and $\tau_v$ were chosen to match the experimental data in the
elastic and viscous-flow regimes, respectively. List of all parameter
values is given in Table I. In (b), the plot of
Ref.~\cite{Johnson}; the curves are artificially shifted to the right to avoid
overlapping.}
\label{fig3}
\end{figure}

When the temperature is fixed, the increasing overshoot with increasing
$\dot{\gamma}$, Fig. \ref{fig3}, is dominated by the exponential in the viscous
stress term. For fixed
material composition (fixed cage
parameters), $\dot{\gamma}$ controls the value of strain $\gamma$ at which the
Newtonian plateau is reached. For high rates of strain $\dot{\gamma}$, the
Newtonian
plateau is shifted to the large strains, and the viscous contribution to the
total
stress is negligible near the yield point $\gamma_{y}$. Since the viscous
stress builds up with increasing $\gamma$, it effectively opposes the decrease
of nonlinear elastic stress due to the cage breakup $n_b(\gamma)$.
Therefore, the stronger 
the viscous stress build-up near the yield point, the less significant is the
nonaffinity-induced stress decrease associated with the overshoot, and the
overshoot itself. If the viscous stress build-up is shifted to large strain,
which happens at high $\dot{\gamma}$, there is no mechanism to compensate the
elastic stress decrease and the overshoot is stronger, which explains why the
amplitude of the overshoot increases with increasing $\dot{\gamma}$.

\begin{table}
\begin{tabular}{|c|c|c|c|c|}
\hline
$K^{[18]}~[\mathrm{GPa}]$ & $\eta^{[18]}~[\mathrm{GPa\cdot s}]$ &
$\tau_c~[\mathrm{s}]$ & $\tau_{v}~[\mathrm{s}]$ & $T^{[14]}~[\mathrm{K}]$\\
\hline
12 & 80 & 1.6 & 2.13 & 523\\
3.6 & 12 & 1.6 & 0.32 & 643\\
1.2 & 4  & 0.6 & 0.12 & 663\\
0.6 & 2  & 0.01 & 0.053 & 685\\
\hline
$K^{[18]}~[\mathrm{GPa}]$ & $\eta^{[18]}~[\mathrm{GPa\cdot s}]$ &
$\tau_c~[\mathrm{s}]$ & $\tau_{v}~[\mathrm{s}]$ &
$\dot{\gamma}^{[18]}~[\mathrm{s}^{-1}]$\\
\hline
3.6 & 80 & 12 & 9.0 & $2\cdot10^{-4}$\\
3.6 & 80 & 12 & 7.3 & $5\cdot10^{-3}$\\
3.6 & 22  & 1.8 & 0.32 & 0.032\\
3.6 & 12  & 1.6 & 0.12 & 0.10\\
\hline
\end{tabular}
\caption{The values of the elastic constant $K$ ($n_b^0=12$), the viscous
constant $\eta$, temperatures, and the strain rates are fixed by 
the characterization of the experimental system for different curves and datasets. The decrease of $K$ and
$\eta$ with temperature are as appropriate for metallic glass, while the
decrease $\eta ( \dot{\gamma})$ reflects the thinning effect in a random
packing at a high shear rate. The values of $\tau_c$ and $\tau_v$ are chosen to
fit the curves in
Figs. \ref{fig2} and \ref{fig3}, the latter changes with $\eta$ as discussed in
the text. The two datasets intersect at the
point $T=643$K, $\dot{\gamma}=0.1$, $K=3.6$, $\eta=12$, $\tau_v=0.16$.}
\end{table}

In Table I we report the values of the physical parameters used for the
plotting of curves. Almost all of these parameters, with the exception of $\tau_{c}$, 
are fixed by the experimental conditions/system, or at least highly constrained.
All the values of the spring constant $K$, of the viscosity $\eta$, and of $T$ and
$\dot{\gamma}$ are fixed by the experiment and taken from \cite{Johnson}. 
{We used the calorimetric glass transition temperature,
$T_{g}=625$ K, determined experimentally for this system~\cite{Peker}. In
reality, there is no such a sharp transition temperature, even in the
calorimetric data, but rather a crossover range which goes from the lower limit
of $625$K up to about $660$K, and the transition temperature also depends
sensitively on the cooling rate. We checked that our results do not change
significantly in the above mentioned temperature transition-range.}. The plateau viscosity decreases
with increasing temperature, an effect common to all dissipating systems -- due
to Arrhenius activation factor always present in $\eta(T)$. The viscous
relaxation time also decreases with $T$, since $\tau_v=\eta / G_R$, and with shear-rate, because
the system is shear-thinning~\cite{Johnson}. Hence, the rheo-physics of the material poses constraints on $\tau_v$
and its dependence on $T$ and $\dot{\gamma}$.
The spring constant $K$ also decreases with $T$ as it is well known
that thermal vibrations reduce the restoring force of the bond~\cite{Born},
whereas its behavior as a function of shear-rate is of less trivial
interpretation. Hence the only non-trivial fitting parameter which can be freely adjusted in our analysis
is the cage relaxation time $\tau_{c}$.

The fact that the
cage relaxation time $\tau_c$ is constant with $T$ but decreases with increasing $\dot{\gamma}$ is
also meaningful. The system is below the glass transition
temperature and the cage parameters should not vary much with temperature in the narrow temperature range under consideration.
The strain rate, instead, which acts like an effective temperature, is varied within a much broader range. In this case
we find a much better fitting if we let the cage relaxation time $\tau_{c}$ decrease significantly upon increasing the strain rate. 
This is another meaningful outcome of our model, because the cage dynamics becomes faster upon increasing the strain rate.

In summary, based on a fundamental atomic-scale stability argument, we derived a theory for the onset of flow in amorphous materials that requires no ad-hoc structural assumption. We believe that this mean-field theory captures essential microscopic ingredients underlying the transition from elastic response to flow: it relates shear-induced configurational nearest neighbor changes directly to mechanical material properties and stress-strain relations.
This coupling leads to an elastic instability at a critical strain $\gamma_y$, when the decreasing local atomic connectivity does no longer allow the lattice free energy to compensate the energy lost to non-affine motions. We showed that this concept provides an atomic-level understanding of the effect of temperature and shear-rate on the emergence and extent of the stress overshoot in metallic glasses.
The presented constitutive stress-strain relation is compact, yet contains all the relevant microscopic physics. It is explicitly presented in the combined Eqs. (\ref{2}) and (\ref{3}). A more elaborate theory, in the future, could explicitly account for the structural heterogeneity of the amorphous structure to induce flow at preferred "weak" locations that - in our framework - have lower coordination. We believe, however, that the simplicity of our mean field theory is also its strength, and, for the first time, allows for clear \textit{microscopic} interpretation of all the involved parameters, and can be more easily implemented for the quantitative analysis of experimental results.


\begin{acknowledgments}
This work was supported by the Theoretical Condensed Matter programme grant from EPSRC.
A.Z. acknowledges financial support by the Ernest Oppenheimer Fellowship at
Cambridge (until 1 June 2014), and by the Technische Universit\"{a}t M\"{u}nchen – Institute for Advanced Study, funded by the German Excellence Initiative and the European Union Seventh Framework Programme under grant agreement n° 291763. 
P.S. acknowledges support by VIDI and VICI fellowships from the
Netherlands Organization for Scientific Research (NWO).
\end{acknowledgments}

\end{document}